\documentclass[twoside,a4paper,11pt]{proceedings}
\usepackage{graphicx}
\usepackage{hyperref}
\usepackage{movie15}
\usepackage{natbib}
\begin{document}
\pagenumbering{arabic}
\pagestyle{myheadings}
\thispagestyle{empty}
\vspace*{-1cm}
%{\flushleft\includegraphics[width=\textwidth,bb=58 650 590 680]{stamp.pdf}}
%{\flushleft\includegraphics[width=\textwidth,viewport=58 650 590 680]{stamp.pdf}}
{\flushleft\includegraphics[width=3cm,viewport=0 -30 200 -20]{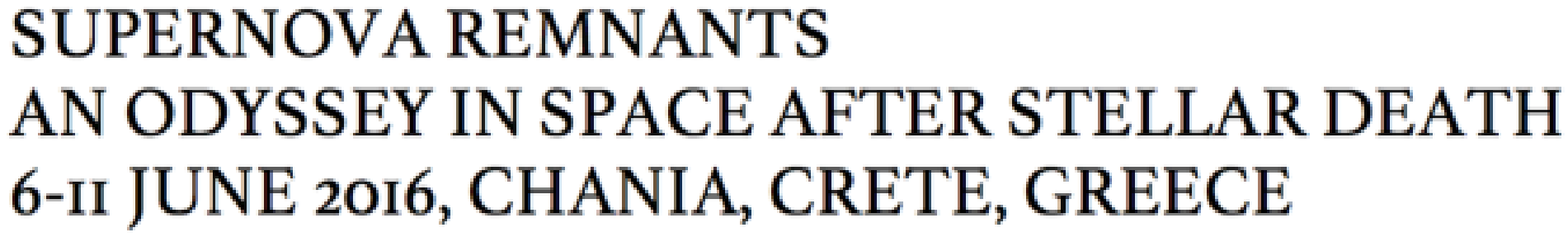}}
\vspace*{0.2cm}
\begin{flushleft}
{\bf {\LARGE
Modelling high-resolution spatially-resolved Supernova Remnant spectra with the Sardinia Radio Telescope \\
%%% TITLE of the paper. 
%Supernova Remnants
}
\vspace*{1cm}
%%% Include here the LIST OF AUTHORS.
Sara Loru $^{1}$$^{,}$$ ^{2}$, 
Alberto Pellizzoni $^{1}$, 
Elise Egron $^{1}$, 
Noemi Iacolina $^{1}$, 
Simona Righini $^{3}$, 
Marco Marongiu $^{1}$, 
Sara Mulas $^{2}$, 
Giulia Murtas $^{2}$, 
Davide Simeone $^{2}$, 
Maura Pilia $^{1}$, 
Matteo Bachetti $^{1}$, 
Alessio Trois $^{1}$, 
Roberto Ricci $^{3}$, 
Andrea Melis $^{1}$
and Raimondo Concu $^{1}$\\
%%% Note that the last author has to be preceeded by an AND.

%Me$^1$,
%You$^2$,
%Somebody Else$^3$
%and Tsikoudias$^4$
%,
%
% Do not delete next few lines
}
\vspace*{0.5cm}
%
%%% AFFILIATIONS LIST.
%%% and the AFFILIATIONS LIST. Note that one affiliation per line.
%%% Add as many affiliations as necessary. 
$^{1}$ INAF-Osservatorio Astronomico di Cagliari, Via della Scienza 5, 09047 Selargius, Italy \\
$^{2}$ Dipartimento di Fisica, Universit\`{a} di Cagliari, SP Monserrato-Sestu, KM 0.7, 09042 Monserrato, Italy \\
$^{3}$ INAF-Istituto di Radioastronomia,Via Gobetti 101, 40129 Bologna, Italy

%
% Do not delete next few lines
\end{flushleft}
% Headings
\markboth{
%%% Type the SHORT version of the paper t
High-resolution spatially-resolved SNR spectra with SRT
}{
%%%  First Author \& Second Author   OR   First-author et al. 
%%%  First Author \& Second Author   OR   First-author et al. if the author list contains three or more authors.
Loru et al.
}
\thispagestyle{empty}
\vspace*{0.4cm}
\begin{minipage}[l]{0.09\textwidth}
\ 
\end{minipage}
\begin{minipage}[r]{0.9\textwidth}
\vspace{1cm}
\section*{Abstract}{\small
%%% Type the ABSTRACT of your paper
%An abstract cannot be made of concrete.
Supernova Remnants (SNRs) exhibit spectra featured by synchrotron radio emission arising from the relativistic electrons, and high-energy emission from both leptonic (Bremsstrahlung and Inverse Compton) and hadronic processes ($\pi^{0}$ mesons decay) which are a direct signature of cosmic rays acceleration. Thanks to radio single-dish imaging observations obtained in three frequency bands ($1.6$, $7$, $22$ GHz) with the Sardinia Radio Telescope (www.srt.inaf.it), we can model different SNR regions separately. Indeed, in order to disentangle interesting and peculiar hadron contributions in the high-energy spectra (gamma-ray band) and better constrain SNRs as cosmic rays emitters, it is crucial to fully constrain lepton contributions first through radio-observed parameters.
In particular, the Bremsstrahlung and Inverse Compton bumps observed in gamma-rays are bounded to synchrotron spectral slope and cut-off in the radio domain. Since these parameters vary for different SNR regions and electron populations, spatially-resolved radio spectra are then required for accurate multi-wavelength modelling.
%Supernova Remnants exhibit spectra featured by synchrotron radio emission arising from the relativistic electrons, and high-energy emission from both leptonic processes like Bremsstrahlung  and Inverse Compton produced by the radio-electrons
%interacting with ambient photons, and hadronic process of $\pi^{0}$ mesons decay which is a direct signature of cosmic rays accelerated in Supernova Remnants. Thanks to radio imaging observations obtained in three frequency bands ($1.4$, $7$, $22 GHz$) with the Sardinia Radio Telescope (www.srt.inaf.it), we can model separately different SNR regions. Indeed, in order to disentangle interesting and peculiar hadron contributions in the high-energy spectra and better constrain SNRs as cosmic rays emitters, it is crucial to fully constrain lepton contributions first through radio-observed parameters.
%In particular, the Bremsstrahlung and Inverse Compton bumps observed in gamma-rays are typically bounded to synchrotron spectral slope and cut-off in the radio domain.Since these parameters vary for different SNR regions and electron populations, spatially-resolved radio spectra are then required for accurate multiwavelength modelling.

\vspace{10mm}
\normalsize}
\end{minipage}
%%% BODY of the paper

\section{Introduction}

%$\,\!$\indent \citet{AbtCard84} happens to be the first paper (in alphabetical order) in my BibTeX file.

%The multiwavelength spectrum of SNRs is typically featured by sinchrotron emission mostly from radio-emitting
%electrons, and high-energy emission arising from bremsstrahlung and Compton processes (IC) produced by the
%radio-electrons interacting with ambient photons, or hadronic emission provided by π 0 mesons decay. 
The long quest for the firm disentanglement among leptonic and hadronic scenarios represents one of
the most important challenges for the high-energy study of SNRs, being directly related to cosmic rays origin
and acceleration models. Accurate radio images of SNRs are typically available at low frequencies (Castelletti G. et al. $2007$ and $2011$; Gao et al. $2011$). On the other hand, multi-wavelength data on SNRs are sparse and spatially-resolved spectra
are  rarely available in the 5-20 GHz range, critical for model assessment, also for the most studied and bright
objects (see Green D.A. $2014$).
%Indeed, recent constraints on cosmic rays emission from SNRs and related models are based on integrated radio fluxes only (no spatially resolved spectra) implying the simplistic ”single-zone” assumption of a single electron population for the whole SNR. 
Deep multi-frequency imaging of the complex SNRs IC443 and W44 with the Sardinia Radio Telescope (SRT) can disentangle different
populations and spectra of radio/gamma-ray-emitting electrons in these SNRs, in order to better address models
and then firmly constrain high-energy emission arising from hadrons. On the other hand, accurate radio spectral imaging allows us to distinguish between shock parameters and different physical processes taking place within SNRs. 
Recent constraints on cosmic rays emission from SNRs and related models (Giuliani et al. $2011$; Ackermann et al. $2013$; Cardillo et al. $2014$) are based on integrated radio fluxes only (no spatially resolved spectra) implying the simplistic ”single-zone” assumption of a single electron population for the whole SNR. 
In the aim of studying the local properties of W44 and IC443, we performed with SRT accurate on-the-fly scans of these SNRs at three
frequencies (L, C, K bands) in order to obtain detailed radio images and
spatial-resolved spectral-slope measurements (synchrotron breaks are possibly
expected in this range). Indeed, spectral index maps provide evidence of a wide physical parameters scatter among different SNR regions.

%: a flat spectrum is observed from the brightest SNR regions at the shock, while steeper spectral index (up to $>$0.8) are observed in fainter cooling regions, disentangling different populations and spectra of radio/gamma-ray-
%emitting electrons in these SNRs.

\section{SRT multi-frequency maps of W44 and IC443}

\begin{figure}[h]
\centering
\includegraphics[scale=0.47]{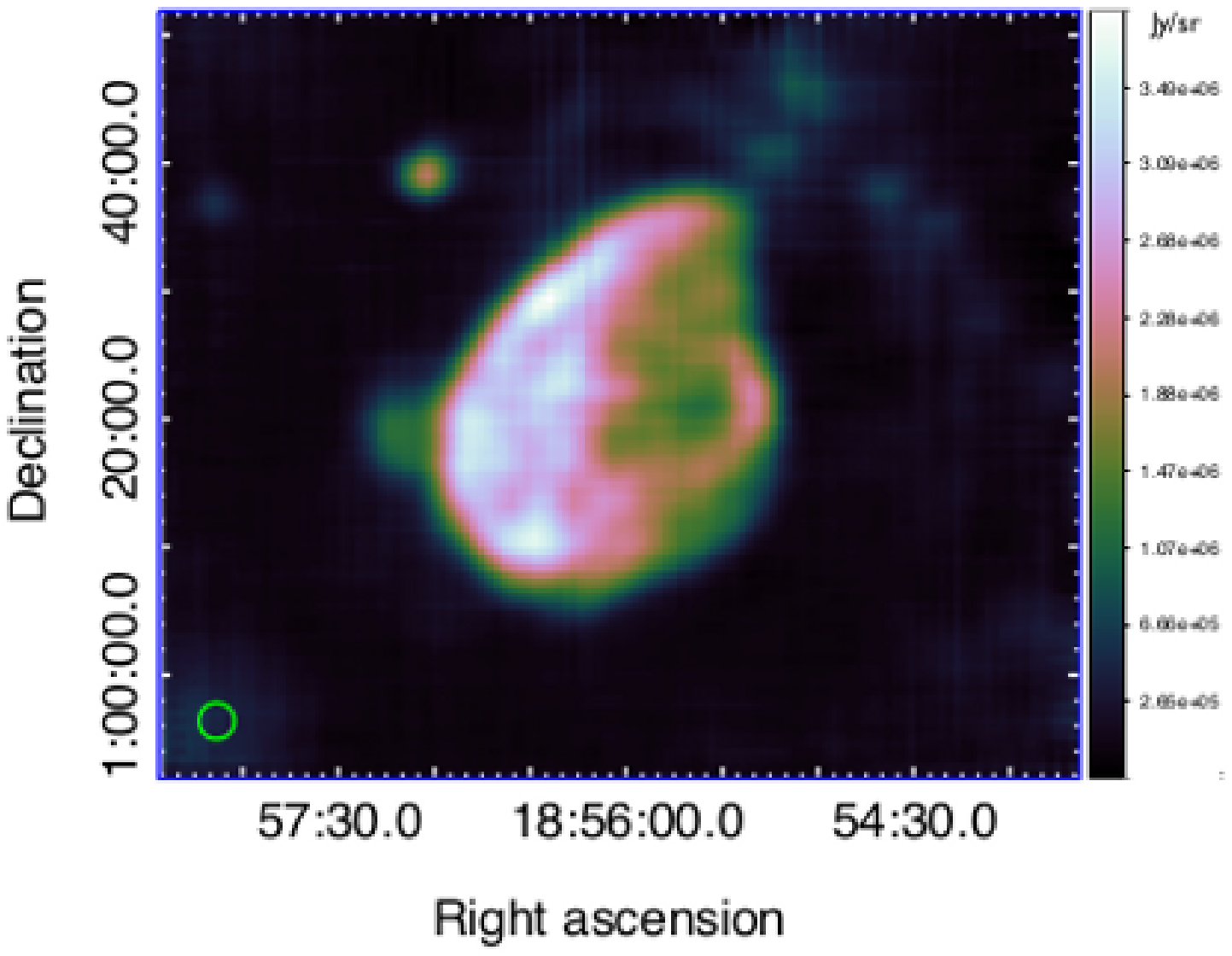} \includegraphics[scale=0.49]{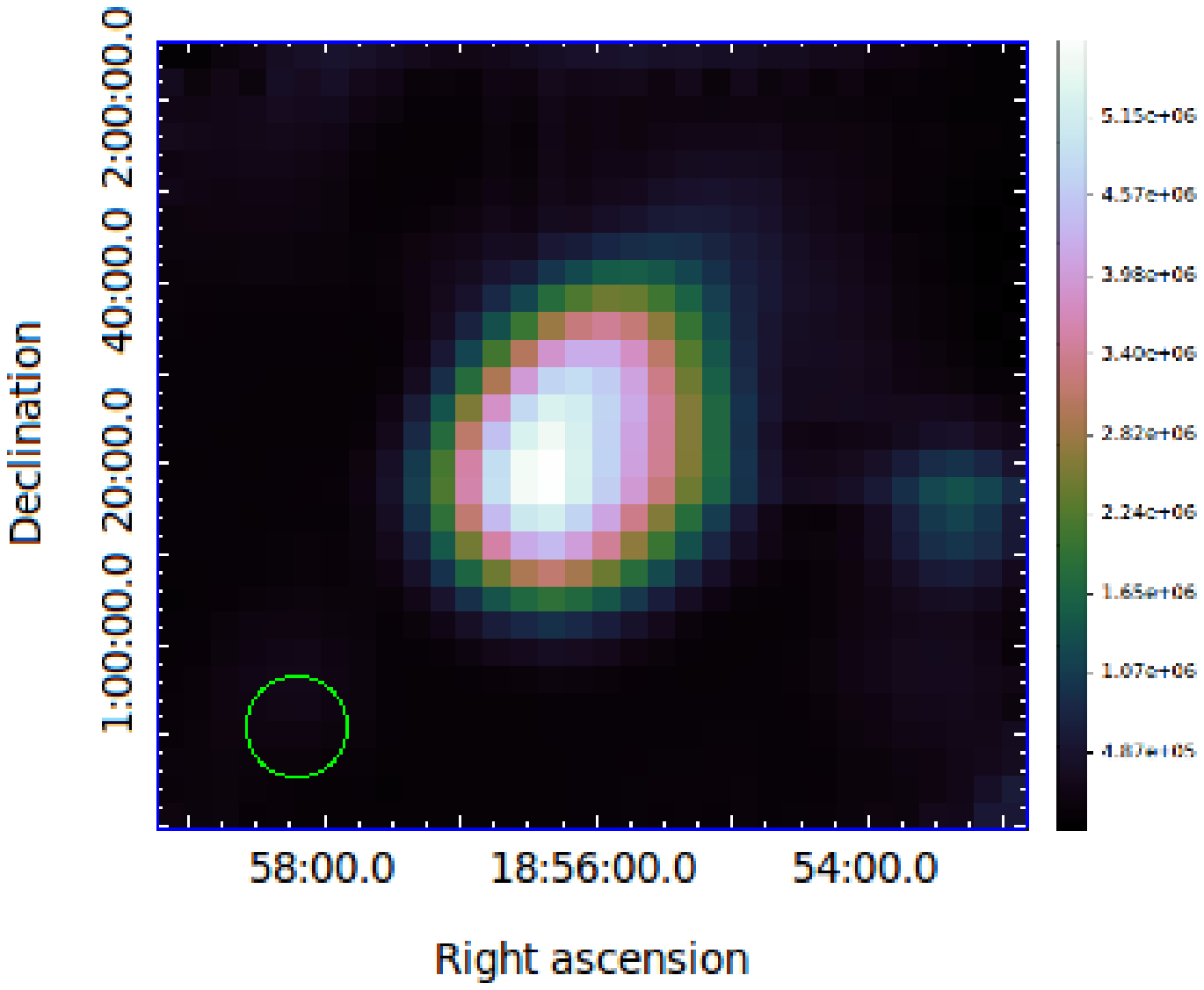}\\
%\vspace{1mm}
\caption{Radio Continuum single-dish maps of SNR W44 performed with the Sardinia Radio Telescope at 7 GHz (C-band, left) and 1.6 GHz (L-band, right). The green circles on the maps indicate the beam size at the observed frequencies (L-band 11.1', C-band 2.7').  }
\label{mappe}
\end{figure}

\begin{figure}[h]
\centering
\includegraphics[scale=0.48]{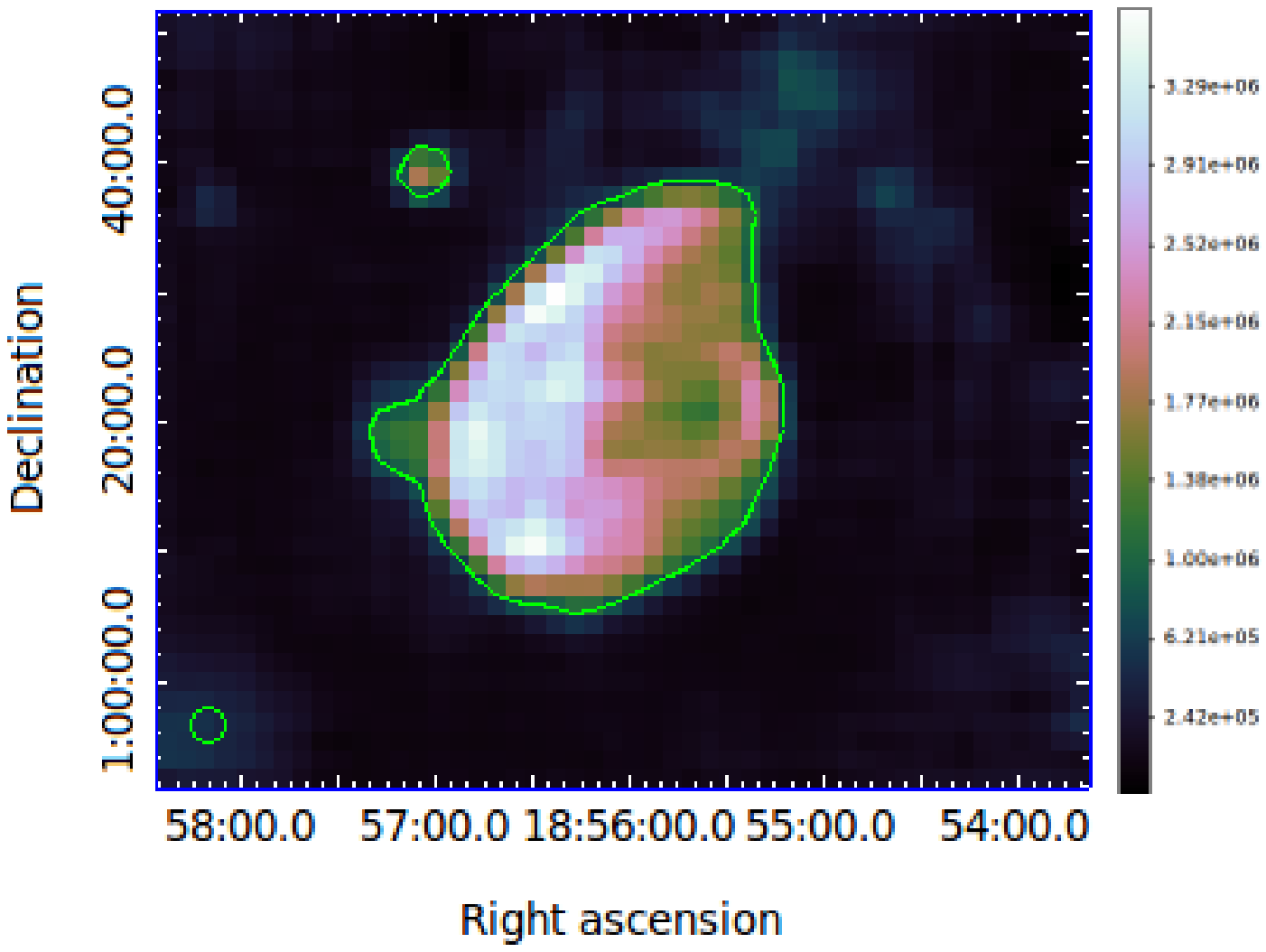} \includegraphics[scale=0.48]{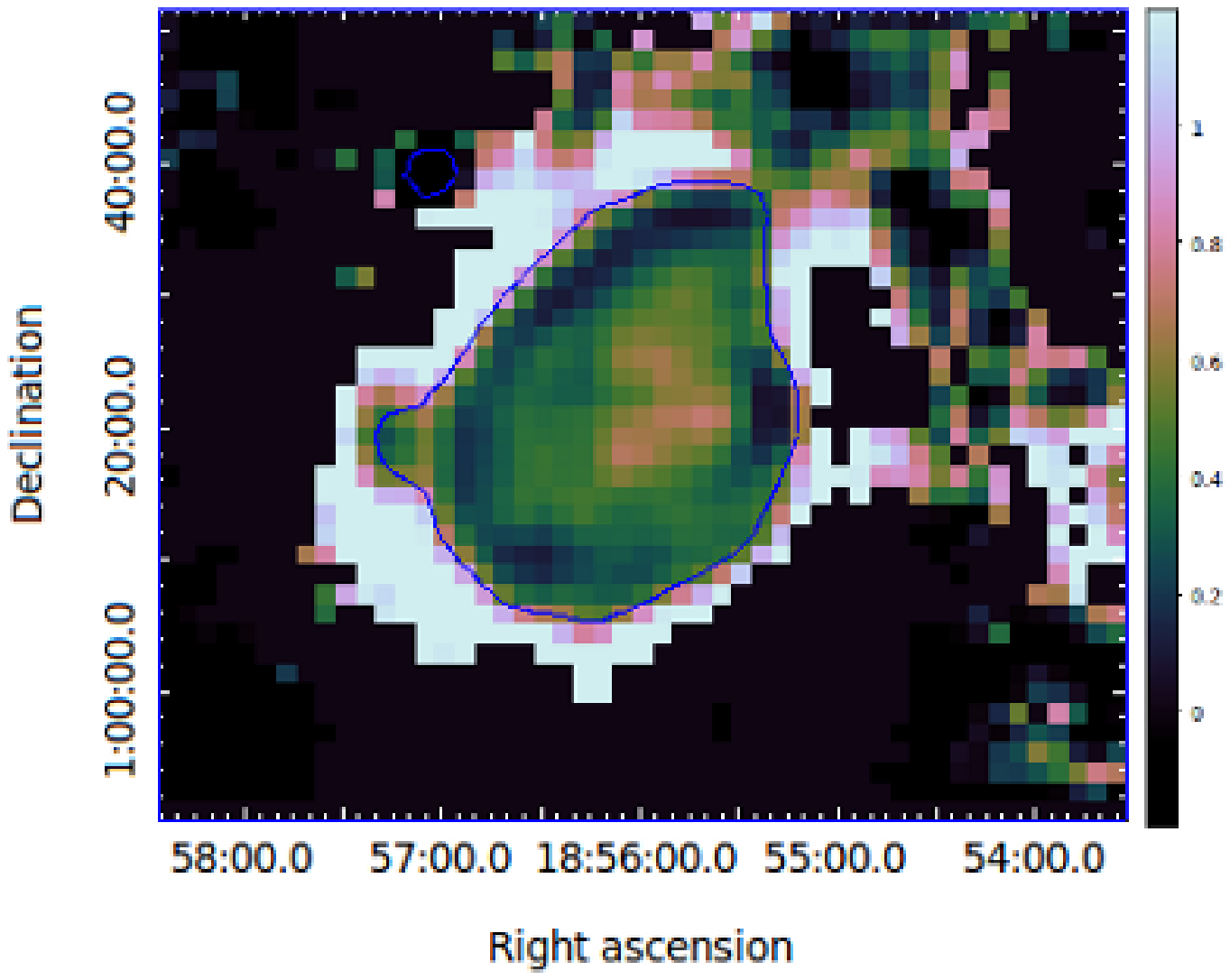}\\
%\vspace{1mm}
\caption{Left: C-band radio continuum map of SNR W44. Right: spectral index map of W44 obtained by coupling 1.6-7.0 GHz data. The contour levels correspond to C-band radio continuum at $\sim$400 mJy beam$^{-1}$. Bright SNR limbs display a flat spectrum while weaker central regions are significantly steeper.}
\label{mappe2}
\end{figure}

In the framework of the Astronomical Validation and Early Science Program with the 64m single-dish Sardinia Radio Telescope,
 we provided for the first time single-dish deep imaging at 7 GHz for IC443 and W44 complexes coupled with spatially-resolved spectra in the range 1.6-7 GHz (Egron et al. these proceedings). In Fig.\ref{mappe} we show the maps of the SNR W44 obtained with SRT at 1.6 GHz and 7 GHz. 
In order to study the spectral index distribution of these sources, we used the spectral index maps obtained by coupling L and C-band images, binning data at low-resolution (1.5' pixel size) and considering only pixels with high source signal in both bands. In Fig.\ref{mappe2} we present the spectral index map of SNR W44 together with the map at 7 GHz for comparison. 
A flat spectrum is observed from the brightest SNR regions at the limbs, while steeper spectral indices (up to $>$0.8) are observed in fainter central regions and halos, disentangling different populations and spectra of radio/gamma-ray-emitting electrons in these SNRs. The possible origin of the observed spectral index scatter is under assessment: it could be due to either non-uniform absorption processes, region-dependent cooling rates or intrinsic shock parameters differences.
In perspective, we will also produce K-band (22 GHz) maps not available so far in literature for IC443 and W44, allowing us to study detailed the morphology of these sources at high radio-frequencies.
Furthermore, coupling K-band maps with L-band and C-band maps, we will search for possible spectral steepening or breaks in selected SNR regions, assessing the high-energy tail of the region-dependent electron distribution.

%\end{column}

%\begin{column}{0.3\textwidth}
%\begin{figure}
%\includegraphics[scale=1]{W44_spectral_index_map_prova2}\\
%\vspace{1mm}
%\caption*{\large \textit{didascalia }}
%\label{mappe}
%\end{figure}
%\end{column}
%\end{columns}

%\section{Conclusion}

%Recent constraints on cosmic rays emission from SNRs and related models are based on integrated radio fluxes only (no spatially resolved spectra) implying the simplistic ”single-zone” assumption of a single electron population for the whole SNR. 
%In the aim of study the local property of W44 and IC443, we performed with SRT accurate on-the-fly scans of these SNRs at three
%frequencies (L, C, K bands) in order to obtain detailed radio images and
%spatial-resolved spectral-slope measurements (synchtron spectral breaks are possibly
%expected in this range). 

% Do not delete the next line
%\small  % Do not delete
%
%%% Comment the following line if you do not have acknowledgments.
%\section*{Acknowledgments}   % Do not delete if you declare acknowledgments
%
%%% ACKNOWLEDGMENTS
%%% ACKNOWLEDGMENTS
%Support for this work was provided by a number of funding agencies that keep giving less money every year due to the financial crisis under the pretext of austerity.

%%% REFERENCES
\section*{References}
\bibliographystyle{aj}
\small
\bibliography{proceedings}
Ackermann et al., $2013$, Science, $339$, $807$;\\
Castelletti G. et al., $2007$, A\&A, $471$, A$537-559$;\\
Castelletti G. et al., $2011$, A\&A, $534$, A$21$;\\
Cardillo et al., $2014$ A\&A,$565$, $74$; \\
Gao X.Y., Han J.L., Reich W. et al. $2011$, A\&A, $5291$, A$159$; \\ 
Giuliani et al., $2011$, ApJ, $742$, L$30$; \\ 
Green D.A. $2014$, A Catalogue of Galactic Supernova Remnants (http://www.mrao.cam.ac.uk/surveys/snrs/) \\
%Sun X. H., Reich P., Reich W. et al. $2011$, A\&A, $536$, A$83$; \\
%SNR1, 2016, MNRAS, 111,222\\
%snr2, 2010, A\&A, 222, 15\\ 

%\begin{columns}
%\begin{column}{0.3\textwidth}

\end{document}